# The Lock Generative Adversarial Network for Medical Waveform Anomaly Detection

Wenjie Xu[a], Scott Dick[a,1]

[a]Department of Electrical and Computer Engineering, University of Alberta, Edmonton T6G 2V4, AB, Canada

**Highlights**

Introduced a workflow to develop the novel Lock Generative Adversarial Networks for the processed clinical waveform data.

Assisted by B-SMOTE and various feature extraction methods, the processed and uniformed input for the network have a portion of contribution on detection accuracy.

The comparison of traditional machine learning methods with state-of-art LGAN networks indicates that our network can achieve competitive and superior results.

Multi-Statistical Significance test was applied on the networks to further validated the practicability and veracity.

**Abstract**

Waveform signal analysis is a complex and important task in medical care. For example, mechanical ventilators are critical life-support machines, but they can cause serious injury to patients if they are out of synchronization with the patient's own breathing reflex. This asynchrony is revealed by the waveforms showing flow and pressure histories. Likewise, electrocardiograms record the electrical activity of a patient's heart as a set of waveforms, and anomalous waveforms can reveal important disease states. In both cases, subtle variations in a complex waveform are important information for patient care; signals which may be missed or mis-interpreted by human caregivers.

We report on the design of a novel Lock Generative Adversarial Network architecture for anomaly detection in raw or summarized medical waveform data. The proposed architecture uses alternating optimization of the generator and discriminator networks to solve the convergence dilemma. Furthermore, the fidelity of the generator network's outputs to the actual distribution of anomalous data is improved via synthetic minority oversampling. We evaluate this new architecture on one ventilator asynchrony dataset, and two electrocardiogram datasets, finding that the performance was either equal or superior to the state-of-the art on all three.

[1]* Corresponding author. *E-mail address:* sdick@ualberta.ca

## 1.  Introduction

Diagnosis and treatment of certain important medical conditions requires the analysis of complex, continuous waveform data. One good example is patient-ventilator asynchrony (PVA), a mismatch between the operation of a mechanical ventilator and the patient's own breathing reflex. PVA is diagnosed by interpretation of the combined flow and pressure waveforms observed from the ventilator by a respiratory therapist [1]. However, these therapists can only spend a relatively brief period with any single patient, and thus can only observe a fairly brief time window of this waveform. In contrast, research shows that different forms of asynchrony manifest at different rates depending on the time of day and other factors. This may be a reason why the prevalence of PVA is not well understood; estimates range from 10% to 85% of ventilated patients. PVA is associated with increased mortality; one study found that an Asynchrony Index greater than 10% was associated with a nearly 5-fold increase in mortality [2]. While exact causality between PVA and mortality has not been established[3], it clearly affects quality of life. Automated PVA detection is designed into many ventilators, but this is only an alarm and can only cover certain predefined condition[4]. However, raising an alarm risk overburdening medical staff. Nurses in an Intensive Care Unit (ICU) can expect many of these alarms, perhaps as many as several hundred per patient, per day. Alarm fatigue – the loss of urgency after constant exposure to frequent, erroneous alarms – is a crucial patient-safety problem, with false-positive alarms seen as perhaps the most damaging to nursing car[5].

In a similar vein, interpretation of electrocardiography (ECG) results plays a prominent role in cardiovascular care. Again, expert interpretation of the ECG waveforms is key; and the time windows that can be analysed are limited. The ECGs taken at a clinic visit are usually on the order of a minute in length. Longer monitoring periods can be recorded via e.g. a Holter monitor but must again be interpreted by a human. The accuracy of those human interpretations is alarmingly low; in a recent meta-study, even trained cardiologists were found to accurately interpret ECGs only about 75% of the time (95% confidence interval 63.2%-86.7%) [1]. In both cases, there is a clear need for technological approaches that can continuously monitor waveform data and identify anomalous patterns even within complex waveforms with high accuracy (i.e. false-positive and false-negative alarms must both be minimized).

We propose a novel Generative Adversarial Network (GAN) as a high-accuracy anomaly detection framework for medical waveform data. Our architecture is designed to overcome the convergence dilemma of traditional GAN architectures using an alternating-optimization approach in which the discriminator is trained individually first. We furthermore employ stratification-based resampling (specifically, the Borderline-SMOTE algorithm[6]) to further reduce false-positive errors. We evaluate this algorithm on one PVA dataset[7], and two ECG datasets[8] [9], finding that LGAN is statistically superior to the existing state-of the-art on the PVA dataset and one of the ECG datasets, and statistically indistinguishable from the state-of-the-art on the remaining one.

The remainder of this paper is organized as follows. In Section 2 we review necessary background and related work on Generative Adversarial Networks, patient-ventilator asynchrony, and ECG interpretation. In Section 3 we propose our Lock-GAN architecture. We discuss our data processing methodology in Section 4, and our evaluation methods and experiment results in Section 5. We close with a summary and discussion of future work in Section 6.

## 2.  Background and Methodology

In this section, we first review Convolutional Neural Networks (CNNs) and Long Short-Term Memory (LSTM) networks[10] [11], and the hybrid ConvLSTM architecture [12]. We then discuss the medical literature on patient-ventilator asynchrony, and electrocardiogram interpretation.

## 2.1. CNN and ConvLSTM Networks

CNNs are a class of feed-forward neural networks whose architecture is fundamentally organized around convolving input signals with trainable, spatially-limited neuron transfer functions. The input signals include (but are not limited to) digital images, and the convolution operation is conceptually organized as a sliding window over the image (similar to an edge detector, or any of several other computer-vision algorithms). This sliding window (referred to as a *kernel*) is applied to every pixel in the image, producing an output *feature map*. Denoting the feature map by $C$, the convolution sum is given by [10]:

$$C[x,y] = \sum_{i=-m/2}^{m/2} \sum_{j=-n/2}^{n/2} K[i,j] I[x-i,y-j] = (I * K)[i,j] \tag{1}$$

at the point $I$ [$x$, $y$] in the input image $I$, for a kernel $K$ whose spatial extent is $m{\times}n$ pixels. This can be implemented as a weighted sum in a neuron assigned to position ($x$, $y$) in an image. The receptive field of that neuron is identically the pixels falling within the kernel window, and the weight of each input (the pixel at position ($i$, $j$) relative to ($x$, $y$)) is given by $K$ [$i$, $j$]. The values $K$ [$i$, $j$] are learned via backpropagation. To ensure that the kernel is everywhere identical for the whole image (and to dramatically reduce the computational effort needed to train the CNN), the weight $K$ [$i$, $j$] is shared between all neurons belonging to that feature map. If we consider a 100×100-pixel image, with a 5×5 kernel at each pixel (and neglecting image boundaries), there are 250,000 weighted inputs in total – but there are only 25 trainable parameters, since the 5×5 kernel is everywhere identical. In a CNN, a single convolutional layer may implement numerous feature maps, by assigning a separate bank of neurons to, and training an independent kernel for, each map [10].

Modern CNNs were introduced by LeCun et al in [13], based on concepts first explored in the Neocognitron [14]. A major insight was that the spatially-limited kernels of a convolution layer would be unable to respond to larger-scale features in the input images. A mechanism is thus needed to widen the spatial extent of the kernels. The modern solution is to intersperse subsampling (or *pooling*) layers between the convolutional layers. Pooling neurons accept the pixel values of a small patch (usually 2×2 pixels) of a feature map and output a single pixel value for the subsampled map. LeCun originally proposed the arithmetic mean of the input values as the pooling function, but the *max*() function is more commonly used today[20,339].

CNNs today are constructed by stacking convolutional and pooling layers into a deep neural network. A number of further innovations (including layer response normalization [15], batch normalization [16], regularization via Dropout [17], residual blocks[18], etc.) have been proposed in the last 25 years. Modern CNNs may be hundreds of layers deep, and outperform all other competitors in many signal-processing tasks (image classification being a prominent example [15]).

The Long Short-Term Memory architecture[11]is a recurrent neural network, meaning that some signals in the network are fed back to previous layers (i.e. closer in the network to the input layer than the output layer). Specifically, LSTM is organized around "memory cells," neural structures that control reading and writing from a self-looping neuron that holds a state across multiple time steps in the network. This is particularly valuable in processing sequence data (time series, etc.) as it allows observations separated by an arbitrary number of time steps to be related to one another. This is a feature of most recurrent networks, but normally the number of time steps must be predetermined so that an appropriate number of delays can be encoded into the network. Such architectures are particularly unwieldy when there are many long-term dependencies present in a sequence. LSTM, by contrast, is able to learn control functions for reading and writing in the memory cells, and thus adapt to whatever delays are present. Numerous memory cells can be assembled into an LSTM, allowing it to model highly complex dependency patterns [11], [19] .

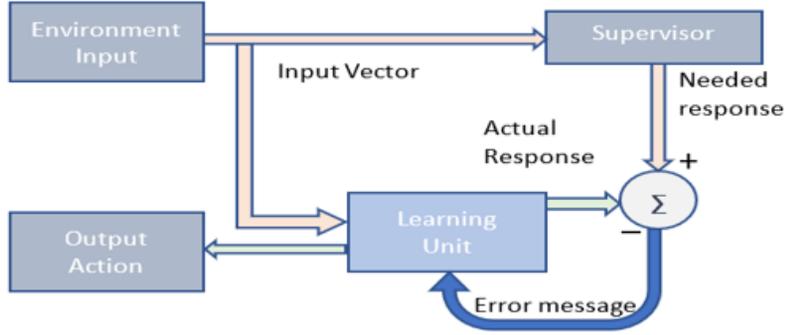

Figure 1: the basic logic hiding in the RNN network, the learning unit is the memory cell which will be updated with each input.

In figure 1, the basic logic of using the memory cell and switch would help to control the information flow and reduce the speed of gradient descent (also known as constant error carousels [20]), which fixed the problem extensively existing in the vanilla RNN architecture [21] [22] [23].

Figure 2 [24] depicts an LSTM memory cell. Several signals are used to control the cell, all of which are generated by neurons trained via backpropagation (technically, backpropagation through time). Denote the stored state at time $t$ by $h_t$. Firstly, the input gate $i_t$ admits a new state into the memory cell to be stored. This gate is controlled by a neuron as follows:

$$i_t = \sigma(W_{xi}x_t + W_{hi}h_{t-1}) \tag{2}$$

where $x_t$ is the input for the current timestamp, $W_{xi}$ is the weight between input lines and $i_t$, and $W_{hi}$ the weights between the input gate and the stored-state neuron $h$. When an input is permitted into the cell, it will be blended with the stored state of the previous time step, $h_{t-1}$.

$$h_t = tanh(W_{xc}x_t + W_{hc}h_{t-1}) \tag{3}$$

This is further modified by the "forget" signal, the "forget" signal clears the memory cell when asserted. and the forget signal of what is inside a LSTM unit, which can be represented by several equations (4-9) below, the major goal of this LSTM learning algorithm is to decide which information are going to be forgot and upkeep. This is the forget gate in equation 4 which a sigmoid function act as a filter.

$$f_t = \sigma(W_{xf}x_t + W_{hf}h_{t-1}) \tag{4}$$

In here, $x_t$ is the input for the current timestamp.

$W_{xi}$ is the weight associated with the input.

$h_{t-1}$ is the previous state passed by the timestamp.

$W_{hf}$ is the matrix weight associated with the previous state.

After the application of the Sigmoid function, $f_t$ would be a number between 0 and 1. In here, 0 means forget everything and 1 means forget nothing. The input gate in equation 5 gives the importance of the new information. The corresponding weight is represented by $W_{xi}$ & $W_{hi}$.

$$i_t = \sigma(W_{xi}x_t + W_{hi}h_{t-1}) \tag{5}$$

The new information that goes throng the cell state would be a function of previous state at the previous timestamp t-1 and the input x at the timestamp t. the activation function tanh in equation 6 is used to make the value of the information to be between -1 and 1. It means that a negative value would make the information deducted from the current cell state and a positive value would make the information added to the cell state.

$$New_t = tanh(W_{xc}x_t + W_{hc}h_{t-1}) \tag{6}$$

One thing that need to mention is that the information should be updated by the following equation 7.

$$c_t = f_t \circ c_{t-1} + New_t \tag{7}$$

where '∘' denotes the Hadamard product. In here, the cell state is $c_{t-1}$  during the current time stamp. Others are the parameters that calculated before.

The output gate in equation 8 is pretty similar with the previous gate which has a sigmoid function and output the values between 0 and 1.

$$o_t = \sigma(W_{xo}x_t + W_{ho}h_{t-1}) \tag{8}$$

To calculate the current hidden state, we use the product of tanh and $o_t$ to update the cell state.

$$h_t = o_t \circ tanh c_t \tag{9}$$

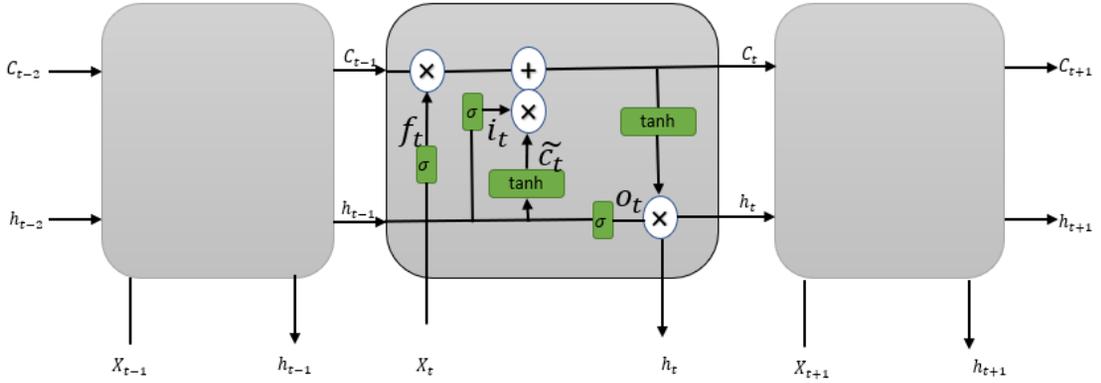

Figure 2: The example of the recurring LSTM layers[24], one of them contains four units interacting within the corresponding input.

As we know, when working with spatial-related data, one of the best approaches nowadays is convolutional architecture. Several filters extract important features when signals pass through convolutional layers. After passing some convolutional layers in sequence, the output is connected to a fully connected dense network. But some data might have the time-related property which can also be well processed by the LSTM network. Then the ConvLSTM [25] used in our encoder-decoder model can be seen as a dimension enhancement layer with a convolution operation in each time step which read a 4D or 5D tensor as input. ConvLSTM is a variant of LSTM (Long Short-Term Memory) containing a convolution operation inside the LSTM cell which is capable of learning long-term dependencies. ConvLSTM replaces matrix multiplication with convolution operation at each gate in the LSTM cell. By doing so, it captures underlying spatial features by convolution operations in multiple-dimensional data which is becoming a more popular model as it is more recent and has shown to provide more inductive priors. The key equation of ConvLSTM input control line changed due to the convolution operation, where ∗ denotes the convolution operator and ⊙ the Hadamard product.

$$i_t = \sigma(W_{xi*}x_t + W_{hi*}h_{t-1} + + W_{ci} \odot C_{t-1} + b_i) \tag{9.1}$$

Corresponding to the LSTM, the forget control line, cell state, output control line, and hidden state equations are represented below with convolutional operation.

$$f_t = \sigma\big(W_{xf} * x_t + W_{hf} * h_{t-1} + + W_{cf} \odot C_{t-1} + b_f\big)$$
$$\tilde{C}_t = tanh(W_{xc} * x_t + W_{hc} * h_{t-1} + b_c)$$
$$C_t = f_t \odot C_{t-1} + i_t \odot \tilde{C}_t \tag{9.2}$$
$$o_t = \sigma(W_{xo} * x_t + W_{ho} * h_{t-1} + + W_{co} \odot C_t + b_o$$
$$h_t = o_t \odot tanh C_t$$

In here, $t$ is the current time step. $x_t$ is the input at the current time step, $h_{t-1}$ is the hidden state from the previous time step. $C_{t-1}$ is the cell state from the previous time step. $h_t$ and $C_t$ are the current hidden state and cell state. $W_{xi}$ $W_{xf}$ $W_{xc}$ $W_{xo}$ are the weights for the input to the input, forget, cell, and output control lines, respectively. $W_{hi}$, $W_{hf}$, $W_{hc}$, $W_{ho}$ are the weights for the hidden state to the input, forget, cell, and output control lines, respectively. $W_{ci}$, $W_{cf}$, $W_{co}$ are the weights for the cell state to the input, forget, and output control lines. $b_i$, $b_f$, $b_c$, $b_o$ are the bias terms for the input, forget, cell, and output control lines, respectively. Symbol * represents the convolution operation. Symbol $\odot$ denotes the Hadamard product. Symbol $\sigma$ represents the sigmoid function, used to calculate the control line activations ranging from 0 and 1.

## 2.2. Patient-Ventilator Asynchrony

Contemporary positive-pressure ventilators support or supplant a patient's breathing reflex by using mechanical pumps to raise the air pressure in the patient's trachea at the onset of a breath. The increased pressure inflates the lungs, delivering oxygen to the patient's bloodstream. The pressure is maintained briefly, then relieved; the natural recoil of the patient's lungs and diaphragm, along with the reduced tracheal pressure, facilitates an exhalation that expels CO2 from the lungs. After a short rest, the cycle repeats[26] . However, this idealized cycle is complicated by the fact that the patient's autonomic breathing reflex is also present; when the two fall out of synchronization (PVA), the result can be a patient struggling against the ventilator, and disrupting the pressure and flow signals that the ventilator's automatic controls rely on. In the worst case, the controller may be confused into raising the tracheal pressures well beyond safe limits – causing lung injury, or even a life-threatening pneumothorax in the worst case [26]. Most modern ventilators include alarm functions that will warn nurses of PVA - but these tend to be simple thresholds, with relatively high error rates (usually biased towards fewer false-negative alarms, at the cost of higher false-positive rates), which are known to lead to alarm fatigue [27].

There are different methods on the dataset of the off-target mechanical ventilation were explored. [26] explored the ERTC, GBC, and MLP classifiers for binary and multi-class breath abnormal value detection which produced a lower accuracy without the ensemble methods. With the remix of these three algorithms, the sensitivity and specificity can be lift but with the compromise of the model complexity. Breath anomaly is often a relatively low chance happened event which give us some difficulty for prepressing the data.[27] sought the method of the synthetic minority oversampling technique (Smote) for the imbalanced data and achieved a uniformed input for the network. Smote which can be extended to regression problems as well by oversampling the continuous-valued variables[28] is also popular among re-sampling methods including the application of epileptic seizures [29], machine Ventilator Asynchrony [26] and other medical and non-medical problems [30], [31].

## 2.3. ECG Interpretation

It's not easy to visually check the ECG waveform abnormal fragments in clinical situation because of its short-time duration and small amplitude. But the detection of certain abnormal is critical for keeping the survival rate of Myocardial infarction (MI) which is caused by the disruption of blood flow by a myocardium segment. The complete disruption of blood flow would result in a heart attack with permanent damage of heart muscle. However, there is no symptoms before the heart attack and the patients don't even aware of the MI until a severe outcome outbreaks. According to American health Association, out of 750000 heart attack Americans almost 1/3 of them have a period attack and 72% of them are quite silent until bad things happened. Even before the heart attack the heart muscles already being destroyed by lacking enough nutrition. As a result, The MI is a critical disease and need the attention of accurate detection system. And the pathway of reducing the high mortality is urgently needed.

But the major problem for manually analysing the ECG signals resulting in the difficulty of detecting and classifying various waveform morphologies in the signal. The characteristics of extremely time-consuming and easy to make errors for categorizing the signals exist even for an experienced cardiologist[32] since cardiac health is constantly monitored by the medical physicians and cardiac practitioners based on the collected ECG leads signals. The machine learning techniques are exploited to handle the problem of manual analysis of the data by detecting the anomalies [33] because deep neural networks in machine learning field are more and more popular in research filed of image and signal processing. A study of ECG waveform data analysis was recently released in[34] using the powerful DNNs. In this paper, the DNNs could achieve similar performance with the single-leads waveform data originally from the 2017 PhysioNet Challenge data[35] and it can reach the highest result with the larger datasets comparing with the clinical cardiologists. But one question that bewilder the authors would be that the DNNs might not powerful when using in a realistic scenario because the physicians usually looking at the 12-lead waveform data, which is the standard practice.

## 3. Lock Generative Adversarial Network Architecture

### 3.1. Relation of GAN, CGAN and LGAN

In 2014, Ian J. Goodfellow et al. proposed the Generative Adversarial Networks (GAN)[36] for generating synthetic fake images to fool the discriminator and eventually the generator can produce the close identical instances compare the ground truth image. The generative models are trying to maximize the joint probability of the P (X, Y) which could be factorized by the Bayes theorem with the presumption that the P(Y). P(X1/Y). P(X2/Y) … P (Xn /Y) are conditionally independence with each other:

$$P(X,Y) = P(Y)P(Y) = P(X)P(X) \qquad (10)$$

While the discriminative models are trying to directly estimate the conditional probability of the target value given the input value p(y|x), which could be transferred by learning the prior distributions p(x|y) and p(y).

$$p(x) = \frac{p(x,y)}{p(x)} = \frac{P(Y)P(Y)}{p(x)} \cong P(Y)P(Y) \qquad (11)$$

The discriminator is sensible for classification problems. These two theories are equivalent in some extent for determining the $\hat{y}$ according to the highest probability $P(X)$. Then GANs have been recently used mainly in creating realistic images, paintings, and video clips. GAN is notorious for training process because of precise hyperparameter requirement, evolving loss landscape and network complexity. Alireza Koochali, et al. [37] explored a method for converting the probabilistic model to determined forecaster which will reduce the complexity of the network. It was also applied in various domain such as Energy industry[38] for data generation process. Ramponi et al.[39] proposed a Time-Conditional Generative Adversarial Network on timestamp information to handle irregularly sampling. Both generator and discriminator in this network are conditional to the time stamp, which learns the relationship of the data and the timeline and suited for the irregularly sampled time series. There are not many applications of GANs being used for detecting medical waveform data as in our case. The main idea, however, should be same — we want to predict breath asynchrony. According to the discriminator in the GAN, the pattern and behaviour of breath should be adjusted and pass them into the generator parameters. Hence, we want to 'generate' data that will have similar distribution as the one we already have from the given features.

GANs can also interpreted as actor-critic like in the reinforcement learning area, but there is no Markov Chains needed and often regarded as producing good samples but lack of good way to quantify this, meaning that there is no standard metrics for evaluating this GAN. A GAN network consists of two models — a Generator (G) and Discriminator (D). The steps in training a GAN are: The Generator is, using random data (noise denoted z), trying to 'generate' data indistinguishable of, or extremely close to, the real data. Its purpose is to learn the distribution of the real data. Then randomly, real, or generated data is fitted into the Discriminator, which acts as a classifier and tries to understand whether the data is coming from the Generator or is the real data. D estimates the (distributions) probabilities of the incoming sample to the real dataset. Then, the losses from G and D are combined and propagated back through the generator. Ergo, the generator's loss depends on both the generator and the discriminator. This is the step that helps the Generator learn about the real data distribution. If the generator does not do a good job at generating a realistic data (having the same distribution), the Discriminator's work will be very easy to distinguish generated from real data sets. Hence, the Discriminator's loss will be very small. Small discriminator loss will result in bigger generator loss (see the equation below for L (D, G)). This makes creating the discriminator a bit tricky, because too good of a discriminator will always result in a huge generator loss, making the generator unable to learn. The process goes on until the Discriminator can no longer distinguish generated from real data. It should mention that the GAN use the Minimax loss for the parameter updating. The Minimax loss is similar with the binary cross-entropy function but with two players involved in. The binary cross-entropy function is:

$$L = -\sum [y_i log \frac{1}{1 + e^{-w^t x_i}} + (1 - y_i) log (1 - \frac{1}{1 + e^{-w^t x_i}})] \quad (12)$$

When we use the estimate $\hat{y}$ to replace the $\frac{1}{1+e^{-w^t x_i}}$ and drop the subscript of $y_i$, the loss function can be abbreviation as:

$$L = -\sum [y log \hat{y} + (1 - y) log log (1 - \hat{y})] \quad (13)$$

Where the y is the ground truth of the class label, $\hat{y}$ is the prediction of the model. The value function for our GAN is the combination of the binary cross-entropy function for the generator and discriminator. In formula 14, the generator will try to minimize the loss while the discriminator will try to maximize it.

$$max_D min_G B(G, D) = E_x \left( log D(x) + E_z \left( lo \, g \left( 1 - D\big(G(z)\big) \right) \right) \right) \quad (14)$$

While $y = 1, L = ln \, ln \, [\hat{y}] = ln \, ln \, [D(x)]$ , When $y = 0, L = ln \, ln \, [\hat{y}] = ln \, ln \, [1 - D\big(G(z)\big)]$. In here, the D(x) is discriminator's predicted probability value for real data. The higher of the value, the realer of the data. G(z) is the generator's output given the noise data n. D(G(n)) is the discriminator's predicted probability value for generated data so called fake data. The discriminator is supervised that give us guidelines for producing good samples.

The algorithm behind of this would be:

Table 1:  The basic idea behind the LGAN and CGAN

| Training loop (for k times training iterations): |
| --- |
| 1: Fix the learning process of Generator:(updating the D) |
|     Start the loop of Discriminator: |
|     ●  Get m real data points $\{x_1, x_2, \ldots, x_m\} \subset X$ |
|     ●  Get m data points sampled from $\{z_1, z_2, \ldots, z_m\} \subset z^d$ from the distribution $\beta$ |
|     ●  Update the gradient descent by $\theta_D$ |
|         $\frac{\partial}{\partial \theta_d} = \frac{1}{m} [ln \, ln \, (D(x)) + ln \, (1 - D(G(z)))]$ |
| 2: Fix the learning of Discriminator: (updating the Condition and GAN) |
|     Start the loop of GAN |

- Get m fake data only
- Update the gradient descent by $\theta_G$

$$\frac{\partial}{\partial \theta_G} = \frac{1}{m}\left[ln\left(1 - D(G(z))\right)\right]$$

The gradient descent method can be extended to standard learning rules with added momentum and/or RMSProp. Despite the difficulty of the convergence, the discriminator could reach the optimal when:

$$D(x) = \frac{p_{data}(x)}{p_g(x) + p_{data}(x)} \tag{15}$$

According to the Jensen–Shannon divergence, we can deduct that the minimal of the generator is reached when the $p_g = p_{data}$, the value function relationship between the real and fake data is shown below:

$$V_{minG} = 2JS(p_g || p_{data}) - 2ln2 \tag{16}$$

The conditional GAN [36]model can be seen as a variant of generative adversarial nets which have extra information $E_{info}$ as condition for both the generator and discriminator. In our case, the information would be the data labels and processed waveform data. In here, we utilize the conditions by adding data labels into the end layer of the discriminator.

The cost function of the conditional adversarial network is shown here:

$$max_D min_G C_{info}(G, D) = E_x\left[logD(x|A) + E_z\left(log\ log\ \left(1 - D\left(G(z|A)\right)\right)\right)\right] \tag{17}$$

Using the mutual information index, we can rewrite the value function as:

$$V_{info}(G, D) = max_D min_G C_{info}(G, D) + \lambda I(x, G(z, x)) \tag{18}$$

In here, the extra info would be $I(x, G(z, x)$ which indicates the mutual information of x and G. It would be 0 if the image x and the target y is totally irrelevant.

While $E$ indicates the expectation with distribution about the variable shown in the script, A is representing the auxiliary information. By conditional GAN, we can learn a conditional probability distribution from our dataset. While the probability distribution of the target values is largely depending on the historical information of the waveformd data, the conditions is hard to be optimal and controlling the generator and discriminator at the same time would resulting in a convergence dilemma [23]. In addition, training GAN and CGAN can be difficult for several failure modes which were usually discussed in some papers, some of the failure modes are shown below: 1: Cannot converge it is often encountered when training, which is resulted from some known and unknown reasons. 2: Losing of gradients: it is frequently happened because the discriminator is often well selected and trained with labels. Due to the powerful regression of the discriminator, the generator is prohibited from improvement resulted from lacking enough information and causing the problem of losing gradients. 3: Mode collapse: It is often happened [40] [41] during training process in which the generator finds out a way to deceive the discriminator and stuck in a local minimum. That certain way is only a few modes in the data distribution which cannot represent the whole modes. As a result, the generator will repeat some results in a certain way and missing some important features. (Info GAN and Label GAN are one kind of the DGAN) So, the Lock GAN is proposed by latch the discriminator network first for several epochs, by simply adjust the generator while locking the discriminator, the training process can be easily controlled.

## 3.2.  *Details on LGAN and its experimental setting*

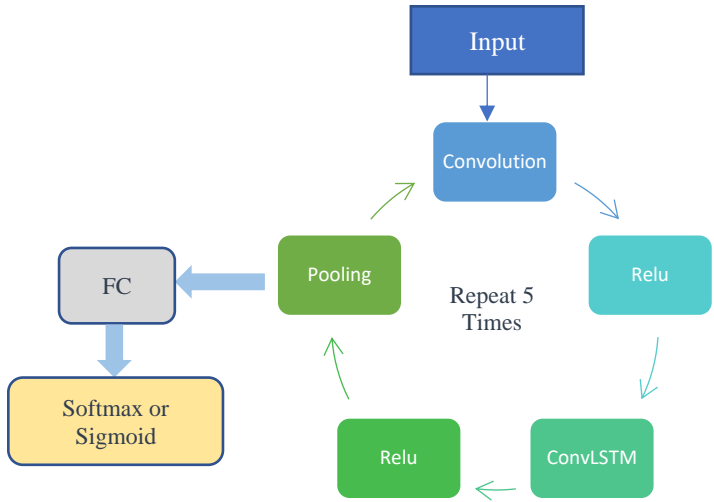

Figure 3: The network for the discriminator is modified as above with a repeating module for 5 times [32]

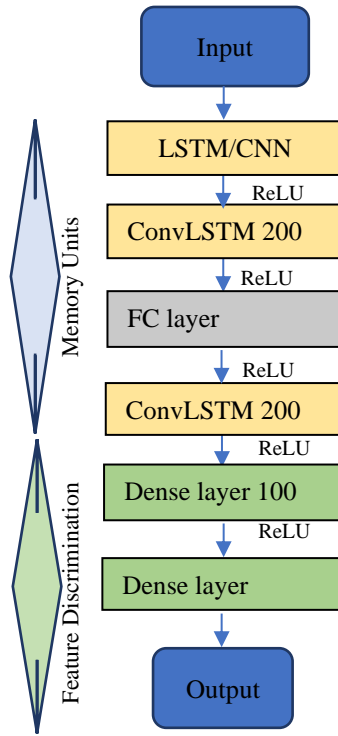

Figure 4: ConvLSTM encoder-decoder model

In following figure 5, we showed the proposed novel LGAN framework for our abnormal data forecasting with figure  and figure 4 as the detailed generator and discriminator model specifications. The procedure can be seen from left to right and from top to bottom. Before feeding the data to the network, the input raw waveform data was processed according to section 4.2, We know that training the GAN is notoriously hard to convergence, so, the advice from the Redford et al [42]and Chollet [43]were explored. Such as, the pooling layer is replaced by the strided convolutional layers. In our model, Figure 4 shows the ConvLSTM encoder-decoder generator model [44] that used in our GAN network, which is a fusion of neural network, Lstm, and ConvLstm. Both of CNN and ConvLstm go through the process of the backpropagation. While we are processing the PVA dataset, the first layer of the model is utilizing the CNN to extract the significant features and the LSTM is used for the datasets to track the inner sequential pattern The hyperparameters are finetuned as Regularization =0.1, momentum = 0.6, and learning rate = 0.005, the weights then updated by gradient descent. The discriminator model shown in figure 3 is a high-performance model can handle the spatiotemporal data, to overcome the drawback of the LSTM-FC or FC-LSTM which used full connections but losing the spatial information [25]of state-to-state or input-to-state transitions, ConvLSTM with pooling layers were used here to capture the inner sequential pattern. The optimal model of the discriminator from the popular backup models [25] [36], [42], [45], [44], [41]which is also testfied by our dataset to achieve the best accuracy. And ConvLSTM determines the future state of a certain cell in the grid by the inputs and past states of its local neighbors which shows better performance compared with FC-LSTM. Also, the number of parameters is reduced largely which is suitable for real-time calculations.  By using the proposed LGAN framework, the space for searching network's architecture would reduce significantly. Without losing the accuracy, this framework offers a sustainable method for adapting highly deterministic model into a probabilistic model and can fully explore the parameters in the GANs.

While in [36], [42], the authors used CNN variants as the major networks to learn the high-quality samples, which is also under our consideration and tested in our dataset with gradient backpropagation and rectified linear activation functions. [42] applied 4 CNN layers for faster calculation and achieved least error rate while testing on the dataset which shows that we can choose less layers to achieve good results instead of sticking on the very deep networks. In [44], the researchers used recurring ConvLSTM layers to represent a narrower range of features and provide some flexibility in learning. So, in our discriminator model, we tried the CNN and the combination of CNN and ConvLstm for testing our dataset with varied number of layers. The best accuracy is achieved by the model in figure 3 which is further used as the lock park in our proposed LGAN model. By confirming our model, we can see that the stacking power of the ConvLSTM with pooling layers from Shi X et al. And Radford A et al 's paper [25], [46]which is used to form the recurring module of our discriminator. The pooling layer following the ConvLSTM is to downsample the feature maps through the summarize of the presence of features. and this is the lock part, which could be an advantage for training and convergence. The true label is given to train the LGAN, which is a condition to make sure the network is well supervised. Once the accurate judger being determined, we will try the generator models for better production of the fake data points which is the imitator of the real data points. The generator nets used a mixture of rectifier linear activations [47] and SoftMax activations. We define a prior on input noise variable which then represent a mapping to data space as G(z;$\theta_g$). The ConvLSTM is used as the main core of our generator, which will give the predicted sequence. A dense layer is added to the ConvLSTM to arrange the output. Based on the popularity and the high performance of the CNN, LSTM, and ConvLstm networks, we run these networks on our dataset and get the best performance which is shown in the results section.

Several cost functions are tried in our model. Such as the discriminator's cost, Minimax. The best performance would be the discriminator's cost which most of the models designed for GANs so far.  Discriminator, J(D) sometime is differed only in terms of the cost used for the generator, J(G).  The cost used for the network is:

$$J^{(D)}\left(\theta^{(D)}\theta^{(G)}\right) = -\frac{1}{2}E_{x\sim P_{data}} \, log \, log \, D(x) \; - \frac{1}{2}E_Z \, log \, log \, \left(1 - D\big(G(z)\big)\right) \qquad (19)$$

While training the classifier with a sigmoid activation layer in the output layer, Equation 21 is the standard cross-entropy cost which will be minimized. The different between the traditional and LGAN network is that the LGAN was trained on two batches with one from the original signal and the other one from the generator where the label is 0 for all examples. We use the LGAN model with 50, 100, and 200 epochs. In terms of the optimization algorithm, we used the Adam optimizer with the best value in 1e-4. Ultimately, we achieved our best results with 100 epochs.

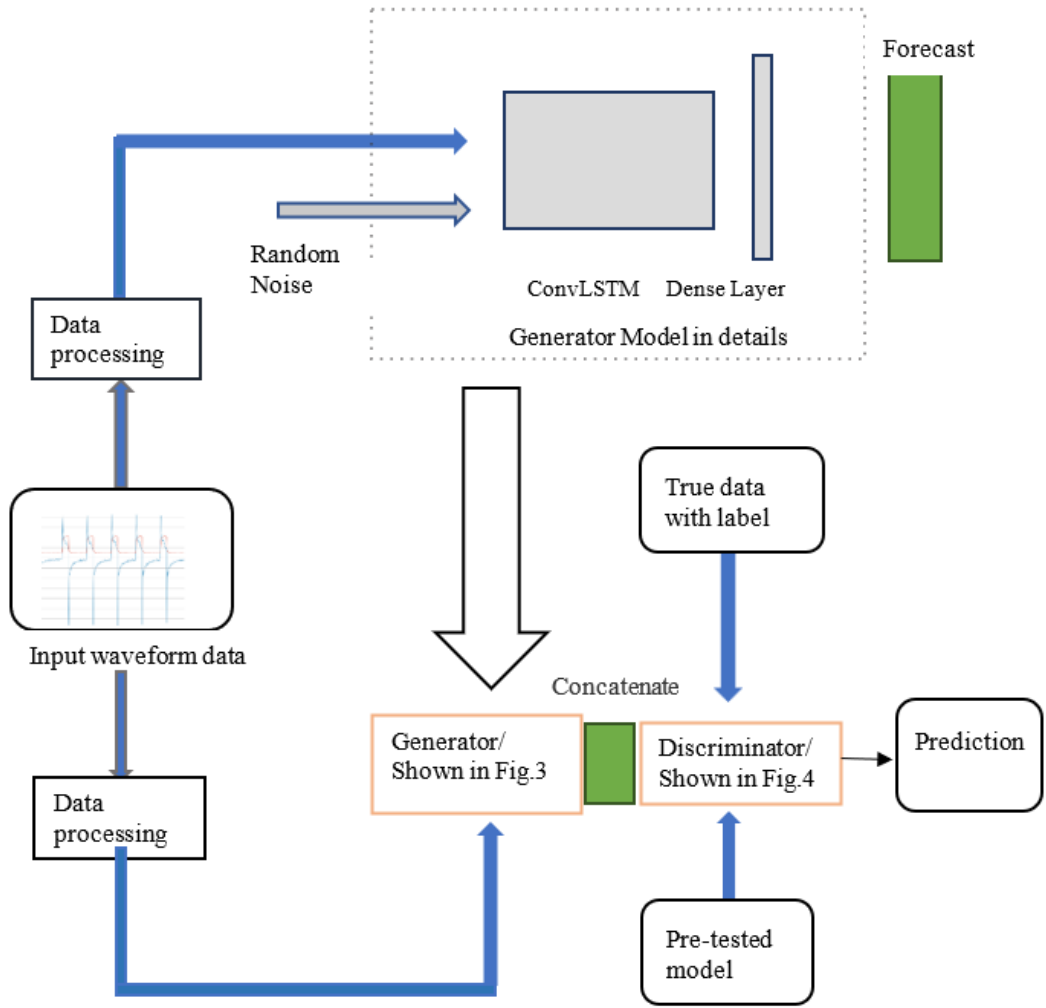

Figure 5: Here is the sketch of the proposed network with the detailed training procedure.

## 4. Data Processing Methodology

### 4.1.    Dataset

The original PVA waveform dataset is outputted from the mechanical ventilators PB840's serial port, with a 50Hz rate of flow, time data, and pressure encoded in ASCII [7]. The total processed data points are 9713 from 37 independent patients with 2998 abnormal points. These files were analysed by identifying regions of interest (ROI) in which the PVA are prevalent, which could ensure that the data contains enough high-quality PVA information. The length of Each ROI is trimmed according to the physician's experience. We use the class category of double trigger asynchrony (DTA), breath stacking asynchrony (BSA), and artifacts included in the ROI, and the total labelled breath length is around 8 hours.

Another two public waveform datasets are the MIT-BIH arrhythmia and PTB ECG database [8], [9]which contains the one-dimensional heartbeats signals. The ECGs in the MIT-BIH Arrhythmia Database has over 4000 Holter recordings which is collected by Beth Israel Hospital Arrhythmia Laboratory. About 60% recordings were directly extracted from the age group from 23 to 89 years and with around 60% are inpatients due to the surgical variations, the same electrode placement cannot be placed in same part in all situations. For example, the lead II in the limb would be replaced by putting it on the chest, which is the standard procedure for the recording.

Del Mar Avionics model 445 two-channel reel-to-reel Holter recorders were used for the recording and the Del Mar Avionics model 660 playback unit was used for the signal's digitization.  The ECGs in PTB Database were obtained in a proprietary compressed format by a non-commercial, PTB prototype recorder with 12 leads. The database contains has 290 subjects with 549 records. Each subject is represented by one to five records.

 These three datasets proved that our proposed network's capability and extensibility because they are large enough for verifying a deep neural network and have many similarities in common waveform dataset area. While the labels for MIT-BIH are ['N': Normal, 'S': Supra-ventricular premature, 'V': Ventricular escape, 'F': Fusion of ventricular and normal, 'Q': Paced • Fusion of paced and normal • Unclassifiable], the labels for the PTB are normal and abnormal which can be tested on our binary classification problem. The Normal and abnormal sample data from ECG are shown in Figure 7.

### 4.2.    Data pre-processing

The raw data from each PVA breath was extracted to metadata, which is a reliable reference for judging the asynchrony and suited for machine learning algorithms. Then we extracted the significant features according to expert knowledge and statistical machine learning methods. The full list of the selected features can be seen in figure 8. Feature selection was conducted using multiple methods including the information gain [46], Chi-square test [48], Fisher's score, correlation coefficient, Support Vector Machine, Recursive Feature Elimination[47], and Principal Component Analysis[49], [50].  We do not want to miss any latent information and try to use all the features to increase the detection accuracy. But the duplicate data would drag the running time and influence the performance of the network in some cases. The computational methods could represent the dataset by ranking the important feature according to some weights. By combining the experts' selection and computational methods, the selected vector was indicated in the binary and multi-class classification section. Such as the figure 8, we carried the mutual information gain for the two separate features according to the discrete label, which give a measurement at a time by producing a non-negative value depending on the correlation of the two random variables. In this information gain process, a non-parametric method based on entropy estimation with a kernel of k-nearest neighbours' distance is used to evaluate the information gain of each variable regarding to the target variable [46]. As we can see, the maxF and minF are  the most irrelevant features, and we just leave these features behind.

The pre-processing methods for ECG data are explored by the studies for processing the signals. such as manually deleting some ambiguous anomaly points or passing the data through some band-width filters[8] . The handcrafted features were used for complicated statistical analysis for the detection and classification problems. but the feature extraction process can be time consuming and lack validity for different patients in various conditions. In the MIT-BIH dataset, the initial beat labels were drafted by the slope sensitive QRS detector which descript all the detected event as a normal heartbeat. Two or more of the cardiologists working independently to annotate the record which is digitized at 360 samples in a second. Third party cardiologists were involved to solve the divergence of the annotation to get the reference label for the computer in each beat.

The cardiologists then added the rhythm labels and abnormal beats. The missed beats, false detections were corrected. The total annotations for the Arrhythmia dataset contains 5 categories and have 109446 samples.

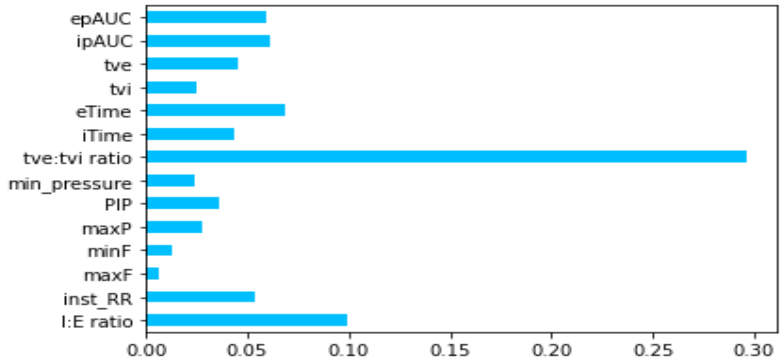

Figure 8: The information gain between different feature groups

In the original PTB ECG dataset, there are 12 leads in total and lead 2 usually gives us a good view of the P wave for a common record of rhythm and strip. So, it is most relevant to our detection problem and was [27]selected with the controlled healthy samples to test our network in the raw waveform format [51], which can be directly used as inputs with minimum pre-processing. in each record, one label was added at the end of the data, which is produced by automatic ECG device or draw by a cardiologist. All records were verified by another cardiologist. for annotating the records, the ECG statements used here is same as the SCP-ECG standard [8]. Finally, the diagnostic classes of subjects are summarized to normal controls and abnormal for binary classification problem which have 14552 samples. For our network purpose, both ECG datasets are fixed dimensions of 144 that are cropped, sampled, and added with zeros if they cannot reach the feature dimensions[9].

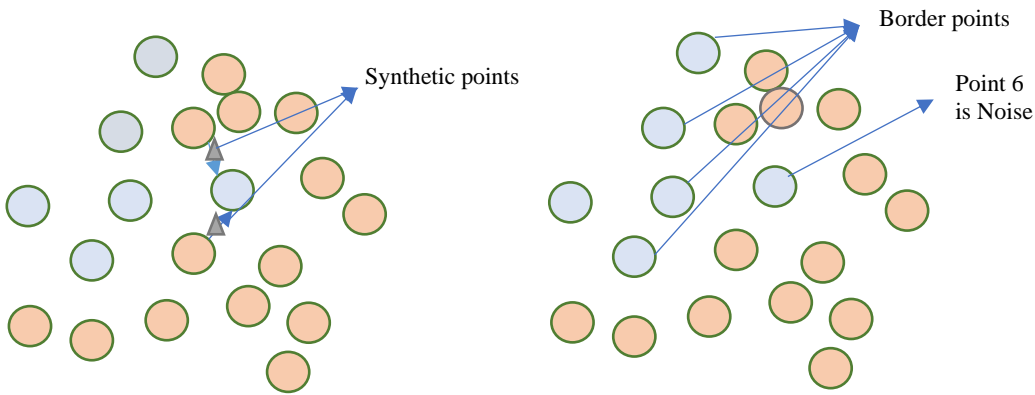

Figure 9: SMOTE and border-Smote selection methods are shown above

Because of the nature of these anomaly points are rare to happen, it brings up the unbalanced problem in our network. To deal with issue, many methods were attempted. Undresampping or downsampling of the majority class is quite common to use in traditional algorithms which select randomly samples from the majority group and make it the same size as the minority class. But this is not an effective method for modern machine learning algorithms to drop a large portion of original data. Oversampling the minority class is a good way sometimes but often results in the overfitting problem with many repetitions of the similar or even same data points. In [52], they proposed the SMOTE which is very popular by helping the production of the minority class. It also points out that the combination with other sample methods might be a good way for imposing the classification performance. In [27], a comparative study was conducted on the Nodule diagnosis and concluded that SMOTE with support vector machine classifier along with the undersampling and oversampling method. While one big problem with that is if the same observations in the minority class appear and outlying the majority class, the Smote would produce a line bridge that confuses the network by mixing up the borders. Like in figure 9, the generated synthetic points between the minority and majority class only make the network hard to converge and sometime would be losing the performance. The Borderline Smote proposed by [53] works the best for our dataset, which is firstly by classifying the minority class samples. And if all or a large portion of the near point are the majority class which is surrounding the minority observation. Then the observation would be defined as noise and be ignored during the synthetic data creation process. Such as in right part of figure 9, the point 6 is defined as a noise which would be an anomaly in the minority class.

In SMOTE, to add 2 times more minority data, we randomly select 2 nearest majority points and produce two more data points between the selected majority and minority points. but in B-SMOTE, while we point and a large portion of normal minority points, the border points are mainly used for the synthetic data generation process. with the help of this technique, our network can increase the accuracy largely.

## 5.  Evaluation methods and experiment results

### 5.1.    Confusion matrix with the K-fold validation:

In this paper, we focus on classifying the asynchrony of the waveform data. The formative problem are binary and multi-classification tasks. To better evaluate the performance on our network, we use the K-fold validation methodology[54] which can avoid the models from picking up the overfit and bise. While a common selectin of K is 10, we choose the 5 as the parameter for training our dataset due to the lacking quantity of the dataset.

The classification performance is usually presented by confusion matrix [55]. A confusion matrix is a M*M square matrix for a M-class classification problem. Each cell in this matrix represents the amount of the observations whether if it is true or false regarding the real label.

In a binary classification problem (in figure 10 right part), if the BSA points are classified into the positive category, the points are so called true positive data. when they were classified into negative category, the points are so called false positive data. Observations of Not-BSA correctly sorted in the Not-BSA category are so called False negative data otherwise, they are true negative data. According to the confusion matrix, we can further get derivations to better descript the performance of specific problem. Such as, in this paper, we choose accuracy, true positive rate (TPR), and false positive rate (FPR) to evaluate the performance.

Accuracy (ACC) is calculated as the total number of all right predictions divided by the overall data points of the dataset. While the 0.0 is the worst case, 1.0 is the best performance. It can also be calculated by $1 - ERR$.

True positive rate (TPR)/ specificity is calculated as the total number of right positive predictions divided by the overall number of positives. While the 0.0 is the worst case, 1.0 is the best performance.

True negative rate (FPR)/specificity is calculated as the total number of right negative predictions divided by the overall negatives. While the 0.0 is the worst negative rate, 1.0 is the best performance.

| Predicted / True class | BSA | DTA | Normal |
|---|---|---|---|
| BSA | A | B | C |
| DTA | D | E | F |
| Normal | G | H | I |

| Predicted / True Class | BSA | Not-BSA |
|---|---|---|
| BSA | TP(A) | FN(B+C) |
| Not-BSA | FP(D+G) | TN(E+F+H+I) |

Figure 10: for the 3-class classification problem, the example of how to derive the TP, FN, FP, FN are shown above.

## 5.2. Binary classification

The PVA and PTB datasets were used for us to test the network for the binary classification as the first step and the PVA class labels are DTA and non-PVA or BSA and non-PVA. The positive class used in here is always setting to the abnormal breath. For BSA, which is caused by respiratory rate is in high frequency and there is not enough time to complete the exhalation between two normal breaths which might be patient or ventilator triggered. For DTA, which occurs due to the shortage of the inspiratory time compared with the patients' wanted i-time and there would trigger another ventilator breath without the exhalation process. Because of the different nature of the PVA, we used the different features for classifying the DTA from BSA. Since the data is imbalanced, we can oversample the minority class by processing methods for our imbalance data. we use Borderline-Smote which use the k-nearest neighbour model to recognize the difficult data points in the minority class. We can oversample difficult examples to gain more detailed resolution. In addition, inspired by the monitoring and inspection [26] of the abnormal data, we use the previous breath metadata for the classification process, which can increase the sensitivity and specificity largely especially for the DTA detection. The previous breath we used is previous one, two, and three. Testified by the GBC, CNN, and ConvLSTM with selected features with previous one, two, and three, we can see the most accurate part of the red line in figure 11 is above the blue dash line and green dash line. we can see that with selected features added with previous two meta data points the best sensitivity and specificity is reached.

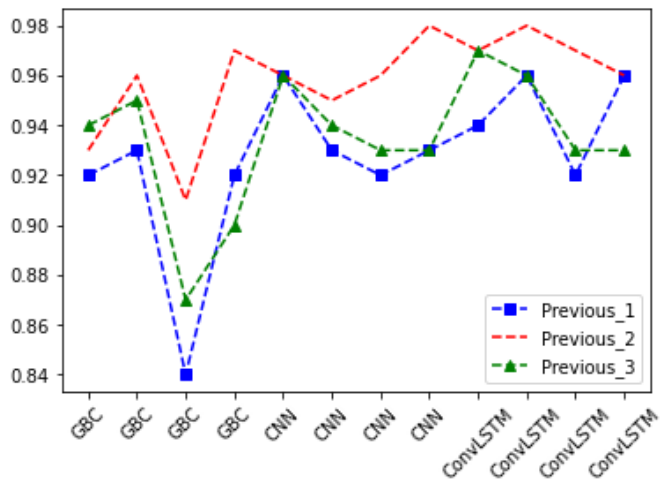

Figure 11: Selected features with previous one, two and three for comparison. For example, for selected features with previous one and same as GBC and CNN algorithm, the four data points for ConvLSTM are BSA sensitivity, BSA specificity, DTA sensitivity, DTA specificity.

The judgement criteria [26] of DTA are quite different from the BSA, so the different features are determined by the feature selection methods. For BSA, the total feature used in here is TVi, TVe, E-time, I-time, peak inspiratory flow, peak expiratory flow, ipAUC, epAUC, and the previous two. For the DTA, the total feature used in here would be I:E ratio, inst_RR, tve:tvi ratio, iTime, eTime, tvi, tve, ipAUC and previous two. The results are shown in table 3-5.

Table 3: The sensitivity and specificity results of different algorithms on the binary classification problem for BSA and normal.

| Algorithm | Sensitivity | Specificity |
|-----------|-------------|-------------|
| LGAN | 0.98 | 0.98 |
| ConvLSTM | 0.97 | 0.98 |
| CNN | 0.96 | 0.95 |
| GBC[26] | 0.93 | 0.96 |
| ERTC [26] | 0.74 | 0.82 |
| RF [26] | 0.96 | 0.97 |

Table 4: The sensitivity and specificity results for different algorithms on the binary classification problem for DTA and normal.

| Algorithm | Sensitivity | Specificity |
|-----------|-------------|-------------|
| LGAN | 0.98 | 0.97 |
| ConvLSTM | 0.97 | 0.96 |
| CNN | 0.96 | 0.98 |
| GBC [26] | 0.91 | 0.97 |
| ERTC[26] | 0.92 | 0.98 |
| RF[26] | 0.98 | 0.84 |

Table 5: The sensitivity and specificity results for different algorithms on the binary classification problem for PTB database.

| Algorithm | Sensitivity | Precision | Accuracy |
|-----------|-------------|-----------|----------|
| LGAN | 0.956 | 0.974 | 96.5 |
| ConvLSTM | 0.942 | 0.952 | 94.8 |
| Acharya et al [56] | --- | --- | 94.7 |
| Safdarian et al [57] | 0.91 | 0.97 | 93.5 |
| Fully Conv [58] | 0.874 | 0.900 | --- |

### 5.3. Multiclassification

First, the imbalance data were pre-processed by B-Smote method to 1:1 ratio. Then we separate the data into 90% and 10% for the training and testing in the following models just same as in the binary problem. The training data is further divided by the 5-fold cross validation method for getting robust performance evaluation. The result in table 6 shows that the proposed LGAN network can achieve a competitive performance for predicting the anomaly points with one group of previous breath metadata and one group of current metadata according to the computational feature selection ranking methods for our results generation with a total feature of 20. Compared with previous studies, we can see the accuracy is increased and the raw confusion matrix is indicated for comparison. In table 7, The pretrained discriminator ConvLSTM reached a relative high accuracy. By testing the proposed network on MIT-BIH arrhythmia dataset, this result shows that the LGAN can get better performance on similar waveform dataset, and we compared the accuracy with existing work, The network achieves the highest result which is 95.6%.

Table 6: The sensitivity and specificity results for different algorithms on the multiclass classification problem.

| Algorithm | Predicted True Class | Accuracy | Non-PVA | DTA | BSA |
|---|---|---|---|---|---|
| LGAN | Non-PVA | 0.973 | 655 | 7 | 9 |
| | DTA | | 15 | 650 | 6 |
| | BSA | | 10 | 7 | 654 |
| ConvLSTM | Non-PVA | 0.969 | 657 | 6 | 8 |
| | DTA | | 7 | 654 | 10 |
| | BSA | | 10 | 21 | 640 |
| CNN | Non-PVA | 0.965 | 660 | 4 | 7 |
| | DTA | | 8 | 640 | 13 |
| | BSA | | 10 | 8 | 643 |
| GBC [26] | Non-PVA | 0.949 | 650 | 8 | 13 |
| | DTA | | 27 | 626 | 18 |
| | BSA | | 25 | 12 | 634 |
| ERTC [26] | Non-PVA | 0.750 | 463 | 108 | 100 |
| | DTA | | 17 | 617 | 37 |
| | BSA | | 86 | 156 | 429 |
| Logistic [7] | Non-PVA | 0.906 | 624 | 20 | 27 |
| | DTA | | 12 | 631 | 28 |
| | BSA | | 42 | 59 | 570 |

Table 7: The accuracy comparison on the MIT-BIH arrhythmia dataset

| Network | Accuracy (%) |
|---|---|
| LGAN | 95.6 |
| CNN | 93.5 |
| Wavelet packet entropy and random forests [59] | 94.61 |
| Deep Transferable Representation[32] | 93.4 |
| Ebrahimzadeh et al. [60] | 95.18 |
| CNN with EMD [61] | 95.98 |

## 6. Statistical Significance test

To know whether there is a significant performance difference between the algorithms, we use ANOVA (known for 3 or more groups testing.) test to find out if experiment outcomes are significant or not. An ANOVA test can tell if the total results are meaningful, but it will not recognize exactly where those differences lie. So, after running an ANOVA (10 replicates in each cell) and found significant results, then we can run post-hoc study Tukey's Tukey Honestly Significant Difference (HSD) to research which groups' means are different, which compares all possible pairs of means giving us to hints if we need to accept the alternate hypothesis or reject the initial null hypothesis. We get 5 accuracy results for each algorithm after running the algorithms, which will be applied on the one-way ANOVA testing first, then we get the p-value which is probably less than the threshold 0.5 with a F-value which means there exists a significant difference between our algorithms or not. For recognizing which algorithm is the winner, the Tukey HSD which used in here compare every mean difference with the other and reject the hypothesis based on the corrected p-value.

Table 8: ONE-WAY ANOVA for the PVA multi-classification

| Source of Variation | Sums of Squares (SS) | Degrees Freedom (df) | Means Squares (MS) | F | PR(>F) |
|---|---|---|---|---|---|
| Between networks | 0.371613 | 5 | 0.074323 | 129 | 8.85e-29 |
| Error (or Residual) | 0.031028 | 54 | 0.000575 | NaN | NaN |

Table 9: ONE-WAY ANOVA for the MIT-BIH dataset

| Source of Variation | Sums of Squares (SS) | Degrees Freedom (df) | Means Squares (MS) | F | PR(>F) |
|---|---|---|---|---|---|
| Between networks | 0.005816 | 5 | 0.001163 | 5.151234 | 0.000618 |
| Error (or Residual) | 0.012194 | 54 | 0.000226 | NaN | NaN |

Table 10: ONE-WAY ANOVA for the PTB database

| Source of Variation | Sums of Squares (SS) | Degrees Freedom (df) | Means Squares (MS) | F | PR(>F) |
|---|---|---|---|---|---|
| Between networks | 0.004567 | 3 | 0.001522 | 8.111778 | 0.000296 |
| Error (or Residual) | 0.006757 | 36 | 0.000188 | NaN | NaN |

Table 11: Multiple Groups Means Comparison of Tukey HSD with family-wise error rate is 0.05for PTB database

| Group pair 1 | Group pair2 | Meandiff | P-adj | Lower | Upper | Reject |
|---|---|---|---|---|---|---|
| ConvLSTM | LGAN | 1.5926 | 0.0016 | 1.0706 | 2.1145 | True |
| LGAN | Achcrya et al. | 1.741 | 0.0012 | 1.2191 | 2.263 | True |
| LGAN | Safdarian et al. | 2.1234 | 0.0034 | 0.9534 | 3.1245 | True |
| Achcrya et al. | Safdarian et al. | -0.2569 | 0.0025 | 1.2358 | 1.6782 | True |
| ConvLSTM | Achcrya et al. | -0.1485 | 0.7242 | -0.6704 | 0.3735 | False |
| ConvLSTM | Safdarian et al. | 1.2459 | 0.0014 | 0.6358 | 1.1468 | True |

In here, Sums of Squares (SS) represents the total variation in the data. It is a measure of the total spread of the data around its mean. The variation Between networks happened because of the differences between the means of the groups. Error (or Residual) is essentially the sum of the squared deviations from their group mean.

Degree of freedom (DF) is a measure of the number of independent values that can vary in this analysis. For example, Degree of freedom between networks is the number of groups minus one.

Mean squares (MS) are the average of the squares of variances, which is calculated by dividing the sum of the squares by the corresponding degrees of the freedom.

F-Value (F) is the statistic calculated by dividing the mean square between the groups by the mean square within the groups. It is used to determine if the means between the groups are significantly different. A higher F-value indicates a greater degree of difference between the group means relative to the variance within the groups.

P-Value (PR(>F)) is the probability of observing an F-value at least as extreme as the one calculated, assuming that there is no difference between the group means. A small p-value of less than 0.05 suggests that there is a statistically significant difference between the groups.In here, group pair 1 and group pair 2 columns are the groups being compared. For the table 8, 9, and 10, the critical values are 2.38, 2.28, 2.86. We can see that there is statistically significant difference between the algorithms regarding each dataset since F_value > 3.5 and P_value < 0.5. From table 11, using Tukey HSD to test differences between groups indicates that overall, there is a statistically significant difference in the different networks. Such as in this Table 11, group1 and group 2 are the control groups which make the comparison. Meandiff is the meaning difference between the two groups. P-adj is the revised p-value for comparing multiple groups. The lower and upper bound are the band of the confidence interval which is at 95% level since alpha = 0.05. Reject means to reject the null hypothesis and there is no difference between the networks. For the dataset PVA and MIT-BIH post-hoc test, we attached them in the appendix due to page limit. From these tables, we can see that LGAN has a significant difference from CNN and ConvLSTm in the dataset of the PVA. But between ConvLSTM and CNN, there is not an obvious significant difference. In MIT-BIH, we can see that CNN with EMD gives the best performance compared with Ebra and LGAN. But compared with our LGAN, there is no significant improvement regarding accuracy. we can see the LGAN achieves the best significance and there is a statistically significant difference between the proposed network with the other networks, which further prove the LGAN can achieve high accuracy with high confidence.

## 7. Conclusion and future work

Despite the intra-patient and inter-patient variation in the waveform data. We proposed a novel Lock GAN framework to train our network with adversarial training. The PVA dataset was extracted from 35 patient's breath data with high prevalence and the 48 half-hour experts labelled ECG data was obtained from 47 subjects. Due to the PVA waveform data are time relevant and high dimensional, we use different statistical machine learning feature selection methods and expert knowledge for feature extraction process. Imbalance data could be an obstacle for achieving robust high accuracy, thus the B-Smote was applied before feeding the data to our network. The research highlights that the waveform data can be processed to categorical data points and feed into the machine learning algorithms like LGAN and ConvLSTM. Our results indicated that it is possible to develop a robust and accurate CGAN-based LGAN detecting model for waveform data prediction according to the processed waveform data. This is tested by three datasets and further verified by the statistical significance test. The application of AI in medicine holds great promise for improving diagnostic accuracy and patient outcomes. However, the potential of these advanced technologies comes with significant ethical and moral implications, particularly regarding explainability. The ability to understand and trust AI decisions is crucial, especially in the sensitive context of healthcare. Therefore, our future work aims to develop methods that enhance the explainability of AI systems, focusing on our LGAN framework. Such as Develop Visualization Tools for LGAN Networks and Analytical Methods for Explainability. Implement advanced visualization techniques could make the inner workings of LGAN networks transparent and understandable. These tools will aim to illustrate how the network processes input data and arrives at conclusions and creating analytical methods that dissect the decision-making process of LGANs would involves breaking down the network's decisions into interpretable components, making it easier to understand how specific features influence the outcomes. In addition, we can also form the problem as a time series problem as a future work to predict the abnormal data in advance and to push the forecast results close to the detection accuracy.

Declaration of competing Interest

The authors have no conflicts of interest to declare.